\documentclass{aa}
\usepackage{graphics, epsfig}
\begin{document}

\title{Pixel lensing observations towards globular clusters}

\author{V. F. Cardone\inst{1} \and M. Cantiello\inst{1,2}}

\offprints{V.F. Cardone, \email{winny@na.infn.it}}

\institute{Dipartimento di Fisica ``E.R. Caianiello'', Universit{\`{a}} di Salerno and INFN, Sezione di Napoli, Gruppo Collegato di Salerno, Via S. Allende, 84081 - Baronissi (Salerno), Italy 
\and INAF - Osservatorio Astronomico di Collurania, via M. Maggini, 64100 - Teramo, Italy}

\date{Received / Accepted }

\abstract{It has been suggested that a monitoring program employing the pixel lensing method to search for microlensing events towards galactic globular clusters may increase the statistics and discriminate among different halo models. Stimulated by this proposal, we evaluate an upper limit to the pixel lensing event rate for such a survey. Four different dark halo models have been considered changing both the flattening and the slope of the mass density profile. The lenses mass function has been modelled as a homogenous power\,-\,law for $\mu \in (\mu_l, \mu_u)$ and both the mass limits and the slope of the mass function have been varied to investigate their effect on the rate. The target globular clusters have been selected in order to minimize the disk contribution to the event rate. We find that a pixel lensing survey towards globular clusters is unable to discriminate among different halo models since the number of detectable events is too small to allow any reliable statistical analysis. 

}

\titlerunning{Pixel lensing towards globular clusters}

\maketitle

\section{Introduction}

The detection of gravitational microlensing effect due to massive astrophysical compact halo objects (MACHOs) is undoubtedly one of the great success stories in astrophysics over the past decade. Since Paczy\'nski's seminal paper (1986), microlensing surveys have begun to monitor millions of sources towards the Magellanic Clouds (\cite{Letal00}, \cite{Aetal01}), the galactic bulge (\cite{Uetal94}), the spiral arms (\cite{Detal01}) and the Andromeda galaxy (\cite{Point01}, \cite{Seb}, Paulin\,-\,Henriksson et al. 2002a,b). The discovery of around 20 candidates towards the LMC and more than a hundred events towards the bulge bears witness to the the potential of microlensing as a tool to investigate both the nature of halo dark matter and the structure of our Galaxy. Despite these successes a number of questions still remain. The optical depth measured towards the galactic bulge is at least a factor two larger than what can be accomodated by theoretical models (\cite{BEBG97}, \cite{Setal99}). The optical depth towards the LMC, measured by the MACHO group (\cite{Aetal01}), is too large by a factor $\sim$\,5 to be accounted for by known stellar populations and too small by the same factor to be due to a halo full of MACHOs. Moreover, the implied MACHO mass range, $(0.1\,-\,1 \ M_{\odot})$, is not easily reconciled with existing constraints on baryonic dark matter candidates (see, e.g., \cite{C94}), even if it has been suggested that MACHOs could be not baryonic at all (\cite{Sazhin}, \cite{Gurevich}). Furthermore, it is still not clear whether the lenses are indeed in the galactic halo since it is also possible that the contribution to the optical depth due to other, so far unknown stellar populations has been seriously underestimated (\cite{B98}, \cite{Zhao99}). In particular, the lensing rate due to stars or MACHOs within the LMC itself may be significant (\cite{Aubetal99}, \cite{Salati99}, \cite{EK00}, \cite{GDG00}).

However, it is important to stress that the interpretation of the results is clouded by our imperfect knowledge of the dark halo structure. In particular, studies show that it is not possible to constrain the halo structure and the lens mass function even with a high number of events if observations are taken in only one line of sight (\cite{MS97}). To overcome this difficulty, the use of globular clusters has been suggested by many authors (\cite{GH97}, \cite{RM98}, \cite{Ursula}). This is possible because globular clusters are distributed throughout the galactic halo, albeit concentrated towards the centre. Therefore, we can probe many lines of sight thus allowing a direct mapping of the variation of the density with position and a determination of the dark halo parameters such as its flattening. Furthermore, if the detailed structure of the halo is lumpy, as suggested by some models of Galaxy formation, monitoring different lines of sight will average the results over such fluctuations in the halo density and thereby alleviate the potential worry about non\,-\,representative lines of sight. Finally, quick calculations (\cite{GH97}) show that self\,-\,lensing may be for sure neglected in globular clusters so that the uncertainties about lens location are greatly reduced. Unfortunately, globular clusters are highly crowded (more than 100 stars\,arcsec$^2$ at the core) thus leading to a small number of resolved stars. This means that classical microlensing (where one monitors resolved stars only) is not useful and pixel lensing (\cite{Crotts92}, \cite{Yannick}) must be considered. In this paper we estimate an upper limit to the event rate for pixel lensing observations towards globular clusters\footnote{A similar study has been done by Rhoads \& Malhotra (1998), but their approach was quite different from the one we will use here. We come back to this point in the conclusions.} in order to investigate whether a monitoring project could discriminate among different halo models. \\

The plan of the paper is as follows. In Sect.\,2 we briefly review some basics on pixel lensing in order to introduce the main formulae needed to estimate an upper limit to the predicted number of microlensing events. The model ingredients we need, regarding both the lenses and the targets, have been described in Sect.\,3, while, in Sect.\,4, we give our results for a selected sample of globular clusters. Conclusions are summarized in Sect.\,5.

\section{Pixel lensing theoretical event rate}

Whilst in classical microlensing one monitors individual sources, in pixel lensing the sources are resolved only when they are lensed. We can therefore only monitor the flux in each detector element rather than the flux from individual sources. If a star is magnified sufficiently due to a lens passing close to the line of sight, then the total flux of the pixel containing the source star (coming from the lensed star, the other nearby unlensed sources and the sky background) will rise significantly above the noise level. The pixel lightcurve has then the classical Paczy\'nski shape (\cite{Pac86}) as the usual microlensing lightcurve of a resolved star. But there is an important difference. Since the star emerges from the background only when it is lensed, the observed pixel lensing event timescale may not be a good tracer of the real Einstein time since we only see the portion of the lensed star lightcurve that emerges above the noise level. This means that the classical formulae used to evaluate the theoretical event rate and to estimate the optical depth from the data do not hold anymore. The first detailed study of pixel lensing was undertaken by Gould (1996) who defined two different regimes\,: a semiclassical regime, where the pixel flux is dominated by the source star and the observable timescale is a good tracer of the Einstein time, and a spike regime, where only high\,-\,magnification events can be detected, without any information on the Einstein timescale. Using Gould's formalism, Han (1996) predicted the pixel lensing event rate towards M31, but this has been obtained assuming a fixed sampling rate and unchanging observing conditions so that it is of limited applicability to ground\,-\,based observations. Here we do not employ Gould's formalism, but follow a different approach (\cite{Point00}) which is more closely related to the pixel method. In order to better illustrate the approach we will use, it is helpful to briefly recall the basic steps of the analysis pipeline adopted for the lensing method (for a detailed description see \cite{LeDu}, \cite{Sebtesi}). The starting point is a set of images of the same target taken on different nights and hence with photometric and seeing conditions changing from one image to another. Before being corrected for seeing variations, the sequence of images must be geometrically and photometrically aligned to the same reference image so that the remaining variations may be due only to seeing effects or microlensing events and variable stars. To minimize the effect of seeing, one uses as base detector element an array of $n$ pixels, which is termed superpixel. The number $n$ is fixed in such a way that the superpixel angular dimensions are larger than the median seeing of the images. For instance, if the median seeing is 1.5\,arcsec and the pixel size is 0.30\,arcsec, we choose $n = 6$ so that the superpixel size is 1.8\,arcsec, larger than the median seeing. Binning the flux in the superpixel reduces the effect of seeing variations, but this is not enough to make them negligible. To this aim, one turns to the pixel method, an empirically derived statistical correction which is applied to each image to match it to the observing characteristics of the reference image\footnote{The pixel method corrects for the different seeing of the images by linearly rescaling the flux of a given superpixel on each image. The scaling coefficients are determined (on an image\,-\,by\,-\,image basis) through a statistical analysis of the differences among the median of the current image and that of the reference image. As a result, the superpixel lightcurve turns out to be flat (within the uncertainties) unless a microlensed star or a variable star is in the field of view of that superpixel. The pixel method has been successfully implemented and has been proven to be an efficient tool to look for microlensing events towards the Andromeda galaxy (Ansari et al. 1997, 1999, Auri\'ere et al. 2001, Calchi Novati et al. 2001, Paulin\,-\,Henriksson et al. 2002a,b). For further details, we refer the reader to the literature (see, e.g., \cite{K98}, \cite{LeDu}, \cite{Sebtesi}.)}. 

Let us consider now a superpixel collecting the photons emitted from a set of stars in the superpixel field of view. Suppose that an unresolved source is undergoing a microlensing event and let us call the source magnification factor $A(t)$. After alignment and seeing corrections, the excess superpixel photon count $\Delta N_{{\rm spx}}$ on an image $i$ obtained at epoch $t_i$ due to a microlensing event is\,:

\begin{equation}
\Delta N_{{\rm spx}}(t_i) = N_{{\rm bl}} [A_{{\rm spx}}(t_i) - 1] = N_{\rm s} f_{\rm see} [A(t_i) - 1] \ ,
\label{eq: spxexcess}
\end{equation}
where $N_{\rm s}$ and $N_{\rm bl}$ are the source and baseline photon counts in the absence of lensing and $f_{\rm see}$ the fraction of the seeing disc contained within the superpixel\footnote{This may be easily estimated as $f_{\rm see} = \pi r_{\rm psf}^2/\Omega_{\rm spx}$ with $r_{\rm psf}$ the radius of the seeing disk (in arcsec). Note that $f_{\rm see} \le 1$ given the way we choose the superpixel dimensions.}. It is important to stress that we are defining $A_{\rm spx}$ as the superpixel count variation factor which acts as the observable counterpart of $A$ since the only quantities which may be detected from observations are $f_{\rm see} N_{\rm s} (A - 1)$ and $N_{\rm bl}$, but not $N_{\rm s}$ and $A$ separately. If we denote the unlensed apparent source magnitude by $m_{\rm s}$ and the surface brightness (in mag$/$arcsec$^{-2}$) of the target galaxy and of the sky respectively as $\mu_{\rm gal}(x,y)$ and $\mu_{\rm sky}$, $N_{\rm s}$ and $N_{\rm bl}$ can be estimated as\,:

\begin{equation}
N_{\rm s} = N_0 {\times} 10^{-\frac{m_{\rm s}}{2.5}} \ ,
\label{eq: nsflux}
\end{equation}

\begin{equation}
N_{\rm bl} = N_0 \ \Omega_{\rm spx} {\times} \left ( 10^{-\frac{\mu_{\rm gal}(x,y)}{2.5}} + 10^{-\frac{\mu_{\rm sky}}{2.5}} \right ) \ ,
\label{nblflux}
\end{equation}
where $\Omega_{\rm spx}$ is the superpixel angular area and $(x, y)$ its coordinates in a rectangular system with origin on the target centre and oriented along its main axes. Note that we are assuming that both $\mu_{\rm gal}$ and $\mu_{\rm sky}$ are constant along the full sequence of images. Because of changing observing conditions, this could not be true, but we may assume that the photometrical aligment has reduced the fluctuations in $\mu_{\rm sky}$ to a negligible level.

The superpixel noise on image $i$ is\,:

\begin{equation}
\sigma_i = max\left \{\sigma_T, \delta_i \sqrt{N_{\rm spx}(t_i)} \right \} \ ,
\label{eq: spxnoise}
\end{equation}
where\,:

\begin{equation}
N_{\rm spx}(t_i) = \Delta N_{\rm spx}(t_i) + N_{\rm bl} \ .
\label{eq: nspx}
\end{equation}
The threshold noise level $\sigma_T$ is determined by the superpixel flux stability, while the correction factor $\delta_i$ takes account of the fact that the pixel method is not photon noise limited. A signal is regarded as being statistically significant if\,:

\begin{displaymath}
\Delta N_{\rm spx}(t_i) \ge 3 \sigma_i \ .
\end{displaymath}
This means that a microlensed source is detected if the magnification is higher than a minimum value which may be computed from Eq.(\ref{eq: spxexcess}) as\,:

\begin{equation}
A_{\rm min}(t_i) = 1 + \frac{3 \sigma_i}{f_{\rm see} N_{\rm s}} \ .
\label{eq: amin}
\end{equation}
A special case occurs when $\sigma_i = \sigma_T$, giving a threshold magnification (\cite{Point00})\,:

\begin{displaymath}
A_T = 1 + 0.0075 \ \frac{N_{\rm bl}}{f_{\rm see} N_{\rm s}} = 
\end{displaymath}
\begin{equation}
\ \ \ \ \ \ =
1 + 0.0075 \ \frac{\Omega_{\rm spx}}{f_{\rm see}} \ \frac{10^{-\frac{\mu_{\rm gal}(x,y)}{2.5}} + 
10^{-\frac{\mu_{\rm sky}}{2.5}}}{10^{-\frac{\rm m_s}{2.5}}} \ .
\label{eq: ath}
\end{equation}
Eq.(\ref{eq: ath}) may be used to define a threshold dimensionless impact parameter $u_T$. The relation between $A$ and $u$ is the same as in classical microlensing (\cite{Pac86})\,:

\begin{equation}
A = \frac{u^2 + 2}{u \sqrt{u^2 + 4}} \ . 
\end{equation}
Inverting this equation to get $u = u(A)$ and using Eq.(\ref{eq: ath}) we can obtain the value of the impact threshold parameter as\,:

\begin{equation}
u_T = \sqrt{2} \left [ \frac{A_T}{\sqrt{A_T^2 - 1}} -1 \right ]^{1/2} \ .
\label{eq: uth}
\end{equation}
The dependence of $u_T$ on $N_{\rm s}$ means that the pixel event rate depends on the source luminosity function $\phi(M)$ with $M$ the absolute magnitude, while the dependence on $N_{\rm bl}$ leads to a spatial variation of $u_T$. Note that $u_T$ depends on the observing filter too. Although in Eq.(\ref{eq: uth}) $N_0$ is dropped out, the dependence on the filter is implicit in the luminosity function since this latter quantity changes according to the observing photometric band.

We can now compute a theoretical pixel lensing event rate at sky coordinates $(x, y)$ as\,:

\begin{equation}
\Gamma_{\rm pl} = \langle u_T \rangle (x, y) \ \Gamma_{\rm c}(x, y) \ ,
\label{eq: gammapl}
\end{equation}
where we have defined\,:

\begin{equation}
\langle u_T \rangle (x, y) = \frac{\int{u_T(M, x, y) \ \phi(M) dM}} {\int{\phi(M) dM}} \ .
\label{eq: utmedio}
\end{equation}
In Eq.(\ref{eq: gammapl}), $\Gamma_{\rm c}(x, y)$ is the classical microlensing event rate which we can estimate as (\cite{RM97}, \cite{J98}) \,:

\begin{equation}
\Gamma_{\rm c}(x, y) = \frac{2}{\pi} \frac{\tau(x, y)}{\langle t_{\rm E} \rangle (x, y)} \ ,
\label{eq: gammac}
\end{equation}
where the optical depth $\tau$ is given by\,:

\begin{equation}
\tau = \frac{4 \pi G D_s^2}{c^2} \int_{0}^{1}{\rho(s) \ s (1 - s) ds} \ ,
\label{eq: tau}
\end{equation}
where $D_s$ is the distance of the target (assumed to be within the galactic dark halo) and $\rho$ the halo mass density expressed as function of $s = D_d/D_s$ (with $D_d$ the distance to the lens). In Eq.(\ref{eq: gammac}), $\langle t_E \rangle$ is the Einstein radius crossing time averaged over the lens mass function and the halo transverse velocity distribution. We use the same procedure described in Han \& Gould (1996) to estimate this latter quantity, but we do not report here the explicit expressions for the sake of shortness.

The pixel lensing theoretical event rate $\Gamma_{\rm pl}$ estimated by Eq.(\ref{eq: gammapl}) must be considered as a strict upper limit to the really observed event rate. Actually, Eq.(\ref{eq: gammapl}) assumes that the observing conditions are unchanged which is not exactly true even if the pixel method partially reduces the variations due to seeing effects and atmospheric conditions. Besides, since one usually has $A_{\rm min} \ge A_T$, Eq.(\ref{eq: gammapl}) also tends to overestimate the true mean pixel lensing cross section. Finally, Eq.(\ref{eq: gammac}) assumes perfect detection efficiency, whilst the efficiency function for pixel lensing surveys is not known at all (but surely much lower than one). These considerations suggest that $\Gamma_{\rm pl}$ cannot be directly compared to observations. Nonetheless, Eq.(\ref{eq: gammapl}) is enough for our preliminary study. Actually, we want to estimate the predicted number of events towards globular cluster to see whether a pixel lensing survey towards these targets may be worthwhile or not and whether the information one can gain is useful in discriminating among halo models.

\section{Model ingredients}

There are different physical ingredients entering Eq.(\ref{eq: gammapl}) that we have to fix in order to estimate the theoretical pixel lensing event rate towards globular clusters. One needs to assume a dark halo mass density profile to compute the optical depth $\tau$ with Eq.(\ref{eq: tau}), while also the transverse velocity distribution and the lens mass function enter the estimate of $\langle t_E \rangle$. On the other hand, the source properties, namely the surface brightness and luminosity function, have to be assigned to estimate the average threshold impact parameter $\langle u_T \rangle$. In this section, we describe our choices for each of these ingredients. As a preliminary remark, we stress that our main aim is to investigate whether a survey looking at globular clusters may be used to discriminate among dark halo models. To this aim, it is useful to observe globular clusters which are far above the galactic plane in order to minimize the disk contribution to the lensing rate. The criteria we employ to select our main targets will be described in Sect. 4. For the time being, let us stress that the only galactic component where we assume the lenses could be is the dark halo so that we have to model only this galactic component.

\subsection{The dark halo}

Modeling the dark halo is not an easy task given the paucity of information we have on the dark matter distribution. Hence, there are different models which fit equally well the same constraints. Furthermore, our position within the Galaxy complicates the measurement of rotation curve which usually gives the strongest constraint on the dark halo, when studying external galaxies. We explore here four different halo models changing both the shape and the slope of the mass density.

The first class of models we consider is the non singular isothermal spheroid with mass density\,:

\begin{equation}
\rho(R,z) = \rho_{\odot} \ \frac{\sqrt{1 - q}}{q \ \arcsin{(\sqrt{1 - q})}}
\ \frac{R_{0}^2 + R_{c}^{2}}{R^2 + z^2/q^2 + R_{c}^{2}} \ ,
\label{eq: rhonsis}
\end{equation}
where $\rho_{\odot}$ is the local dark matter density, $q$ the halo flattening, $R_0$ the distance of the Sun to the galactic centre and $R_c$ the core radius. The {\it local} parameters are fixed as\,:

\begin{displaymath}
\rho_{\odot} = 0.008 \ M_{\odot}/pc^{-3} \ , \ R_0 = 8.0 \ {\rm kpc} \ ,
\end{displaymath}
whilst we explore models with three different values of $q = 0.5, 0.75, 1.0$. Finally, the halo core radius is fixed by the following relation\,:

\begin{equation}
R_c = \left [ \frac{v_{c}^{2}(\infty)}{4 \pi G \rho_{\odot}} \ 
\frac{\arcsin{(\sqrt{1 - q})}}{\sqrt{1 - q}} - R_{0}^{2} \right ]^{1/2} \ ,
\label{eq: rc}
\end{equation}
where $v_c(\infty)$ is the asymptotic value of the flat rotation curve fixed to be $200$\,km$/s$. The mass density is all we need to evaluate the optical depth towards a target with galactocentric coordinates $(D_s, l, b)$\,:

\begin{displaymath}
\tau(D_s, l, b) = \frac{4 \pi G D_s^2}{c^2} \ \frac{(R_0^2 + R_c^2) \ 
\sqrt{1 - q}}{q \ \arcsin{(\sqrt{1 - q})}} \ \rho_{\odot} \ {\times}
\end{displaymath}
\begin{equation}
\ \ \ \ \ \ \ \ \ \ \ \ \ \ \ 
\int_{0}^{1}{\frac{s (1 - s)}{A_{\rm iso} s^2 + B_{\rm iso} s + C_{\rm iso}} \ ds}
\label{eq: tauiso}
\end{equation}
with\,:

\begin{displaymath}
A_{\rm iso} = D_s^2 \ (\cos^2{b} + q^{-2} \sin^2{b}) \ ,
\end{displaymath}
\begin{displaymath}
B_{\rm iso} = -2 R_0 D_s \cos{b} \cos{l} \ ,
\end{displaymath}
\begin{displaymath}
C_{\rm iso} = R_0^2 + R_c^2 \ .
\end{displaymath}
To estimate the average time duration $\langle t_E \rangle$, we need to know the transverse velocity distribution of the lenses. This should be obtained in a dynamically self\,-\,consistent way starting from the mass density to recover the model distribution function in the configuration space and then integrating over the spatial coordinates and the velocity parallel to the line of sight. This is not a difficult task for spherical models (see, e.g., \cite{BT87}), but the result could be complicated to obtain for spheroidal models. However, we can follow the usual approach assuming that the transverse velocity distribution for the isothermal models is isotropic and maxwellian\,:

\begin{equation}
f(v_{\perp}) \ dv_{\perp} \propto \frac{2 v_{\perp}}{v_{H}^2} \ 
\exp{\left (\frac{v_{\perp}^{2}}{v_{H}^2} \right )} \ dv_{\perp}
\label{eq: fviso}
\end{equation}
with $v_H = 210 \ {\rm km/s}$.

The second model we consider is the so called {\it truncation flat} model (hereinafter TF model) which has been successfully used to reproduce the dynamics of the Magellanic Stream (\cite{Lin}) and the motions of the Milky Way's satellite galaxies (\cite{WE99}). The mass density is spherically symmetric and it is given as\,:

\begin{equation}
\rho(r) = \frac{M_{tot}}{4 \pi r^2} \ \frac{r_{s}^{2}}{(r^2 + r_{s}^{2})^{3/2}} \ ,
\label{eq: rhotf}
\end{equation}
where $M_{tot}$ the total mass of the halo and $r_s$ a scale radius. We fix these parameters as found by Wilkinson \& Evans (1999)\,:

\begin{displaymath}
M_{tot} = 1.9 {\times} 10^{12} \ M_{\odot} \ , \ r_s = 170 \ {\rm kpc} \ .
\end{displaymath}
Inserting Eq.(\ref{eq: rhotf}) into Eq.(\ref{eq: tau}) we get the following expression for the optical depth\,:

\begin{equation}
\tau(D_s, l, b) = \frac{G M_{tot} D_{s}^{2} r_{s}^{2}}{c^2}
\int_{0}^{1}{\frac{s (1-s)}{P(s) \ [P(s) + r_{s}^{2}]^{3/2}} \ ds}
\label{eq: tautf}
\end{equation}
with\,:

\begin{displaymath}
P(s) = A_{\rm TF} s^2 + B_{\rm TF} s + C_{\rm TF}
\end{displaymath}
with\,:

\begin{displaymath}
A_{\rm TF} = D_s^2 \ , \ B_{\rm TF} = -2 R_0 D_s \cos{b} \cos{l} \ , \  C_{\rm TF} = R_0^2 \ .
\end{displaymath}
Wilkinson \& Evans (1999) have computed the distribution function for the TF model. We have numerically integrated their expression in the case of isotropy in the velocity space and interpolated the results to get the following approximated expression for the transverse velocity distribution of the model we are using\,:

\begin{equation}
f(v_{\perp}) \ dv_{\perp} \propto v_{\perp} \exp{\left [ -1.151 \left ( 
\frac{v_{\perp}}{0.5984 \ v_s} \right )^{1.0379} \right ]} \ dv_{\perp}
\label{eq: fvtf}
\end{equation}
with\,:

\begin{displaymath}
v_s = \sqrt{G M_{tot}/r_s} \simeq 219 \ {\rm km/s} \ .
\end{displaymath}
The last ingredient we need to fully characterize our halo models is the local mass function (hereinafter MF) of the lenses\footnote{Following Han \& Gould (1996), we are implicitely assuming that the MF is homogenous, i.e. $dn/d\mu = \rho(R,z)/\rho_{\odot} {\times} \psi(\mu)$, so that we have only to give the local MF $\psi(\mu)$ (see also \cite{Lukas}, \cite{CdRM01}).}. In the usual approach, one assumes that all the MACHOs have the same mass so that the MF is a $\delta$\,-\,Dirac centered on the common mass $\mu$ (expressed in solar units). However, this distribution is not realistic and can lead to seriously wrong results since the average time duration is quite sensitive to the MF. A first step towards a more realistic approach is to use a power\,-\,law MF\,:

\begin{equation}
\psi(\mu) \propto \mu^{-\alpha} \ .
\label{eq: psimf}
\end{equation}
The slope $\alpha$ and the lower and upper mass limits of the lenses $(\mu_l, \mu_u)$ are poorly constrained. However, comparing the data towards LMC with the theoretical expectations (also taking into account the detection efficiency and the self\,-\,lensing effects), Cardone et al. (2001) have estimated the slope $\alpha$ of the local MF using the same spheroidal models we are employing here. Even if the model parameters used are slightly different, we will use their results in our computations. They did not consider the TF model in their analysis so that we will fix the parameters of the MF for the TF model to the values they have obtained for their isothermal spherical models. Although quite arbitrary, this choice is enough for our aims and more realistic than the usual common mass hypothesis. Furthermore, we will consider two values for $\mu_l = 0.001, 0.01$, whilst the upper mass limit is fixed to be $\mu_u = 1.0$ (\cite{CdRM01}). Table\,1 summarizes the main properties of the different models considered and introduce the labelling we will use in the following.

\begin{table}
\begin{center}
\begin{tabular}{|c|c|c|c|c|c|}
\hline
Id & $\rho(R, z)$ & $f(v_{\perp})$ & $q$ & $\alpha_0$ & $(\alpha_l, \alpha_u)$ \\
\hline \hline
I50a & Eq.(\ref{eq: rhonsis}) & Eq.(\ref{eq: fviso}) & 0.50 & 1.10 & (0.50, 1.48) \\ 
I75a & Eq.(\ref{eq: rhonsis}) & Eq.(\ref{eq: fviso}) & 0.75 & 1.17 & (0.80, 1.50) \\ 
I100a & Eq.(\ref{eq: rhonsis}) & Eq.(\ref{eq: fviso}) & 1.00 & 1.21 & (0.84, 1.56) \\
TFa & Eq.(\ref{eq: rhotf}) & Eq.(\ref{eq: fvtf}) & 1.00 & 1.21 & (0.84, 1.56) \\
\hline
I50b & Eq.(\ref{eq: rhonsis}) & Eq.(\ref{eq: fviso}) & 0.50 & 1.12 & (0.27, 1.69) \\ 
I75b & Eq.(\ref{eq: rhonsis}) & Eq.(\ref{eq: fviso}) & 0.75 & 1.34 & (0.77, 1.89) \\ 
I100b & Eq.(\ref{eq: rhonsis}) & Eq.(\ref{eq: fviso}) & 1.00 & 1.34 & (0.77, 1.90) \\ 
TFb & Eq.(\ref{eq: rhotf}) & Eq.(\ref{eq: fvtf}) & 1.00 & 1.34 & (0.77, 1.90) \\
\hline
\end{tabular}
\end{center}
\caption{Labelling and main properties of the models considered in our analysis. The first half of the table refers to models with $(\mu_l, \mu_u) = (0.001, 1.0)$, whilst the second one is for models with $(\mu_l, \mu_u) = (0.01, 1.0)$. The lower and upper limit of the range used for $\alpha$ are $(\alpha_l, \alpha_u)$ with $\alpha_0$ as {\it best fit} value (\cite{CdRM01}).}
\end{table}

\subsection{The globular cluster model}

In order to use Eq.(\ref{eq: gammapl}) to evaluate the pixel lensing theoretical event rate, we need to give also the surface brightness of the target. The globular cluster surface brightness is very well fitted by the King model (King 1962, 1966)\,:

\begin{displaymath}
\mu_{\rm GC}(R) = \mu_0 \ + \ 5 \log{\left [ 1 - \frac{1}{(1 + r_{t}^{2}/r_{c}^{2})^{1/2}} \right ]} \ -
\end{displaymath}
\begin{equation}
\ \ \ \ \ \ \ \ \ \ \ \ \
5 \log{\left [\frac{1}{(1 + R^{2}/r_{c}^{2})^{1/2}} - \frac{1}{(1 + r_{t}^{2}/r_{c}^{2})^{1/2}} \right ]}
\label{eq: mugc}
\end{equation}
with $\mu_0$ the central surface brightness (in mag$/$arcsec$^2$), $r_t$ the tidal radius and $r_c$ the core radius. These parameters will be fixed as reported in the catalogue\footnote{We have used the 1999 version.} of galactic globular clusters compiled by Harris (1996) which can be found under {\tt http://physun.physics.mcmaster.ca/Globular.html}.

A key ingredient in the evaluation of the event rate by Eq.(\ref{eq: gammapl}) is the target luminosity function (hereinafter LF) which enters the determination of the average threshold impact parameter $\langle u_T \rangle$. However, the measurement of the globular cluster LF is a quite complicated task because of the severe crowding. Typically, the use of the {\it Hubble Space Telescope} allows one to resolve a sufficient number of stars, but the brightest ones cannot be observed because of saturation of the WFPC instrument. A possible solution is to combine ground\,-\,based observations of the brightest (and resolved) stars in the cluster with the faintest star observed with the HST. Unfortunately, this method has been used to determine the LF only for four of the globular clusters we have selected, namely NGC\,5272 (\cite{M3lf}), NGC\,6254 (\cite{M10lf}), NGC\,6341 and NGC\,7078 (\cite{M92lf}). Furthermore, the measured LF is incomplete at the lower end, while a correct estimation of $\langle u_T \rangle$ invokes the knowledge of the LF over the full brightness range. Thus, we choose to adopt a LF derived via numerical simulations rather than an experimental one. The use of theoretically derived LF allows us to take into account the presence of stars in all evolutionary stages: from the faint and richly populated end of the Main Sequence, to the bright and fast Asymptotic Giant Branch (AGB) and Post-AGB phases. The LF used in this work have been calculated implementing the evolutionary code developed for the synthesis of stellar population by Brocato et al. (1999, 2000). The results of this code have been largely tested on the young LMC clusters, and on the old galactic globular clusters showing a very good agreement. 

We briefly report here the main assumptions and ingredients of the code.  All the computations refer to simple stellar populations, i.e. single\,-\,burst stellar systems with homogeneous chemical composition. Once the age $T$ and the chemical composition $(Z, Y)$ of the stellar population are fixed, the mass spectrum of the evolving stars in the population is obtained assuming a Scalo\,-\,like Initial Mass Function, randomly populated through a Monte Carlo method. The physical quantities associated to each evolving mass (luminosity, effective temperature, surface gravity, etc.) are derived adopting the scenario of stellar evolution of the Teramo\,-\,Pisa\,-\,Roma group (see Cassisi et al. 1998 and references therein). Hence, to relate the theoretical quantities $\log{L/L_{\odot}}$ and $\log{T_{eff}}$ with the observational {\it colours} and {\it magnitudes}, we assumed the atmospheres models by Lejeune et al. (1997). For a detailed description of the complete computational procedures of the code we remind to the quoted Brocato et al. papers. 

Being interested to the galactic globular clusters, in the following we refer to LF obtained from simulations of simple stellar populations with a {\it fiducial} age $T = 15$\,Gyr and for three different chemical compositions: $Z = 10^{-4}, Y = 0.23$ (metal\,-\,poor populations); $Z = 10^{-3}, Y = 0.23$; $Z = 10^{-2}, Y = 0.25$ (metal\,-\,rich populations).

\section{Selection of globular clusters and results}

The catalogue of globular clusters compiled by Harris (1996) contains almost 150 systems and many of them are distributed in the galactic halo. In order to best probe the dark halo structure, it is important to reduce as much as possible the contribution of the other galactic components to the event rate so that we can safely assume that the events that will be (eventually) detected are due to halo lenses. To this aim, we will consider as targets only those globular clusters which are far above the disk plane, i.e. with $|b| > 10^o$. This constraint reduces also the bulge contribution. However, to completely exclude the possibility that the line of sight goes through the bulge, we further impose $X < 6$\,kpc, where $X$ is the $x$\,coordinate of the cluster in a rectangular system centered on the Sun with the $x$\,-\,axis positively directed towards the galactic centre. We also ask that our targets reside inside the galactic halo so that we will not consider globular clusters with $D_{\rm GC} > 100$\,kpc, with $D_{\rm GC}$ the distance of the cluster to the galactic centre. Using these selection criteria we have compiled a sample of potentially interesting targets among which we only select those clusters which are observable from the TT1 telescope\footnote{The TT1 telescope is situated in Toppo di Castelgrande (Potenza, Italy) with coordinates ($40.6^o$\,N, $15.8^o$\,E) and will be used by the SLOTT\,-\,AGAPE collaboration to monitor the Andromeda galaxy searching for microlensing events by the pixel lensing method. Details on the TT1 may be found at the web site {\tt http://www.na.astro.it/oacmedia/tt1\_site/index.html}, whilst further informations on the SLOTT\,-\,AGAPE project are available at {\tt http://www.na.astro.it/slag/slott.html}.}. For these systems, we have evaluated the theoretical pixel lensing event rate using Eq.(\ref{eq: gammapl}) assuming that MACHOs made up 20\% of the halo dark matter (\cite{Letal00}, \cite{Aetal01}). 

\begin{table}
\begin{center}
\begin{tabular}{|c|c|c|c|c|}
\hline
Name & $\tau_{I50}$ & $\tau_{I75}$ & $\tau_{I100}$ & $\tau_{TF}$ \\
\hline \hline
NGC 288 & 1.22 & 1.06 & 0.91 & 1.48 \\ 
NGC 1904 & 2.27 & 1.69 & 1.34 & 1.95 \\ 
NGC 2419 & 12.3 & 9.74 & 7.86 & 9.48 \\ 
NGC 4147 & 3.30 & 3.39 & 3.31 & 4.96 \\ 
NGC 5024 & 3.23 & 3.38 & 3.36 & 5.17 \\ 
NGC 5053 & 2.93 & 3.01 & 2.95 & 4.61 \\ 
NGC 5272 & 1.75 & 1.61 & 1.46 & 2.38 \\ 
NGC 5466 & 3.14 & 3.23 & 3.18 & 4.99 \\ 
NGC 5904 & 1.61 & 1.39 & 1.23 & 2.56 \\ 
NGC 6171 & 1.66 & 1.31 & 1.09 & 2.97 \\ 
NGC 6205 & 1.51 & 1.22 & 1.02 & 1.83 \\ 
NGC 6218 & 0.89 & 0.67 & 0.55 & 1.22 \\ 
NGC 6254 & 0.72 & 0.54 & 0.44 & 0.95 \\ 
NGC 6341 & 1.67 & 1.31 & 1.09 & 1.87 \\ 
NGC 6366 & 0.48 & 0.35 & 0.28 & 0.58 \\ 
NGC 7078 & 2.64 & 2.07 & 1.70 & 2.97 \\ 
NGC 7089 & 3.06 & 2.63 & 2.27 & 4.10 \\ 
Pal 13 & 5.79 & 5.38 & 4.81 & 6.94 \\
\hline
\end{tabular}
\end{center}
\caption{Optical depth (in units of $10^{-8}$) towards the globular cluster centre for the selected targets. Note that we have dropped the label a or b from the model id since $\tau$ does not depend on the MF.}
\end{table}

\begin{table*}
\begin{center}
\begin{tabular}{|c|c|c|c|c|c|c|c|c|}
\hline
Name & $\Gamma_{\rm c}(I50a)$ & $\Gamma_{\rm c}(I75a)$ & $\Gamma_{\rm c}(I100a)$ & $\Gamma_{\rm c}(TFa)$ & $\Gamma_{\rm c}(I50b)$ & $\Gamma_{\rm c}(I75b)$ & $\Gamma_{\rm c}(I100b)$ & $\Gamma_{\rm c}(TFb)$ \\
\hline \hline
NGC 288 & $2.4_{-1.1}^{+1.4}$ & $2.1_{-0.6}^{+1.1}$ & $2.0_{-0.9}^{+1.7}$ & $3.7_{-1.7}^{+3.2}$ & $1.6_{-0.6}^{+0.9}$ & $1.5_{-0.6}^{+0.8}$ & $1.3_{-0.5}^{+0.7}$ & $2.4_{-0.9}^{+1.3}$ \\ 
NGC 1904 & $3.6_{-1.7}^{+2.2}$ & $2.8_{-1.3}^{+1.5}$ & $2.4_{-1.1}^{+2.1}$ & $4.1_{-1.9}^{+3.6}$ & $2.5_{-0.9}^{+1.4}$ & $2.0_{-0.7}^{+1.0}$ & $1.6_{-0.6}^{+0.8}$ & $2.7_{-1.0}^{+1.4}$ \\ 
NGC 2419 & $10.1_{-4.8}^{+6.0}$ & $8.3_{-3.8}^{+4.5}$ & $7.2_{-3.4}^{+6.3}$ & $10.8_{-5.1}^{+9.4}$ & $6.8_{-2.6}^{+3.7}$ & $5.9_{-2.2}^{+3.0}$ & $4.7_{-1.8}^{+2.5}$ & $7.0_{-2.6}^{+3.7}$ \\ 
NGC 4147 & $4.6_{-2.2}^{+2.7}$ & $4.7_{-2.1}^{+2.6}$ & $4.9_{-2.3}^{+4.3}$ & $8.6_{-4.0}^{+7.5}$ & $3.1_{-1.2}^{+1.7}$ & $3.3_{-1.2}^{+1.7}$ & $3.2_{-1.2}^{+1.7}$ & $5.6_{-2.1}^{+2.9}$ \\ 
NGC 5024 & $4.6_{-2.2}^{+2.7}$ & $4.8_{-2.2}^{+2.6}$ & $5.1_{-2.4}^{+4.5}$ & $9.1_{-4.2}^{+8.0}$ & $3.1_{-1.2}^{+1.7}$ & $3.4_{-1.3}^{+1.7}$ & $3.4_{-1.2}^{+1.7}$ & $6.0_{-2.2}^{+3.1}$ \\
NGC 5053 & $4.0_{-1.9}^{+2.4}$ & $4.2_{-1.9}^{+2.3}$ & $4.5_{-2.1}^{+3.9}$ & $8.5_{-3.9}^{+7.4}$ & $2.7_{-1.0}^{+1.5}$ & $3.0_{-1.1}^{+1.5}$ & $2.9_{-1.1}^{+1.5}$ & $5.6_{-2.1}^{+2.9}$ \\ 
NGC 5272 & $3.0_{-1.4}^{+1.8}$ & $2.9_{-1.3}^{+1.6}$ & $2.8_{-1.3}^{+2.5}$ & $5.4_{-2.5}^{+4.7}$ & $2.0_{-0.8}^{+1.1}$ & $2.0_{-0.8}^{+1.1}$ & $1.8_{-0.7}^{+1.0}$ & $3.5_{-1.3}^{+1.8}$ \\ 
NGC 5466 & $4.5_{-2.2}^{+2.7}$ & $4.7_{-2.1}^{+2.5}$ & $4.9_{-2.3}^{+4.3}$ & $9.0_{-4.2}^{+7.9}$ & $3.1_{-1.2}^{+1.7}$ & $3.3_{-1.2}^{+1.7}$ & $3.2_{-1.2}^{+1.7}$ & $5.9_{-2.2}^{+3.1}$ \\ 
NGC 5904 & $3.2_{-1.6}^{+1.9}$ & $2.9_{-1.3}^{+1.6}$ & $2.8_{-1.3}^{+2.4}$ & $6.6_{-3.1}^{+5.8}$ & $2.2_{-0.8}^{+1.2}$ & $2.1_{-0.8}^{+1.1}$ & $1.8_{-0.4}^{+0.9}$ & $4.4_{-1.6}^{+2.3}$ \\ 
NGC 6171 & $3.6_{-1.7}^{+2.2}$ & $2.9_{-1.3}^{+1.6}$ & $2.7_{-1.2}^{+2.3}$ & $8.6_{-4.0}^{+7.5}$ & $2.4_{-0.9}^{+1.3}$ & $2.1_{-0.8}^{+1.1}$ & $1.7_{-0.6}^{+0.9}$ & $5.6_{-2.1}^{+2.9}$ \\ 
NGC 6205 & $3.0_{-1.4}^{+1.8}$ & $2.5_{-1.1}^{+1.4}$ & $2.3_{-1.1}^{+2.0}$ & $4.7_{-2.2}^{+4.1}$ & $2.0_{-0.8}^{+1.1}$ & $1.8_{-0.6}^{+0.9}$ & $1.5_{-0.6}^{+0.8}$ & $3.1_{-1.1}^{+1.6}$ \\ 
NGC 6218 & $2.2_{-1.1}^{+1.3}$ & $1.7_{-0.8}^{+0.9}$ & $1.5_{-0.7}^{+1.3}$ & $4.0_{-1.8}^{+3.5}$ & $1.5_{-0.6}^{+0.8}$ & $1.2_{-0.5}^{+0.6}$ & $1.0_{-0.4}^{+0.5}$ & $2.6_{-1.0}^{+1.4}$ \\ 
NGC 6254 & $1.9_{-0.9}^{+1.1}$ & $1.5_{-0.7}^{+0.8}$ & $1.3_{-0.6}^{+1.1}$ & $3.3_{-1.5}^{+2.9}$ & $1.3_{-0.5}^{+0.7}$ & $1.0_{-0.4}^{+0.5}$ & $0.8_{-0.3}^{+0.4}$ & $2.1_{-0.8}^{+1.1}$ \\ 
NGC 6341 & $3.3_{-1.6}^{+2.0}$ & $2.7_{-1.2}^{+1.4}$ & $2.4_{-1.1}^{+2.1}$ & $4.7_{-2.2}^{+4.1}$ & $2.2_{-0.8}^{+1.2}$ & $1.9_{-0.7}^{+1.0}$ & $1.5_{-0.6}^{+0.8}$ & $3.1_{-1.1}^{+1.6}$ \\ 
NGC 6366 & $1.4_{-0.7}^{+0.8}$ & $1.0_{-0.5}^{+0.6}$ & $0.9_{-0.4}^{+0.8}$ & $2.2_{-1.0}^{+1.9}$ & $0.9_{-0.4}^{+0.5}$ & $0.7_{-0.3}^{+0.4}$ & $0.6_{-0.2}^{+0.3}$ & $1.4_{-0.5}^{+0.7}$ \\ 
NGC 7078 & $4.6_{-2.2}^{+2.7}$ & $3.7_{-1.7}^{+2.0}$ & $3.3_{-1.5}^{+2.9}$ & $6.7_{-3.1}^{+5.8}$ & $3.1_{-1.2}^{+1.7}$ & $2.6_{-1.0}^{+1.3}$ & $2.1_{-0.8}^{+1.1}$ & $4.4_{-1.6}^{+2.3}$ \\ 
NGC 7089 & $5.0_{-2.4}^{+3.0}$ & $4.5_{-2.0}^{+2.4}$ & $4.1_{-2.0}^{+3.6}$ & $8.7_{-4.0}^{+7.6}$ & $3.4_{-1.3}^{+1.9}$ & $3.2_{-1.2}^{+1.6}$ & $2.7_{-1.0}^{+1.4}$ & $5.7_{-2.1}^{+3.0}$ \\ 
Pal 13 & $7.1_{-3.4}^{+4.3}$ & $6.6_{-2.7}^{+3.5}$ & $6.2_{-2.9}^{+5.4}$ & $10.7_{-5.0}^{+9.4}$ & $4.8_{-1.8}^{+2.6}$ & $4.6_{-1.7}^{+2.4}$ & $4.1_{-1.5}^{+2.1}$ & $7.0_{-2.6}^{+3.7}$ \\
\hline \hline
Total & $72.0_{-34.6}^{+60.5}$ & $62.2_{-27.0}^{+52.2}$ & $61.1_{-28.4}^{+53.6}$ & $119.0_{-55.3}^{+104.0}$ & $48.9_{-18.6}^{+26.6}$ & $45.6_{-16.9}^{+23.4}$ & $40.1_{-14.9}^{+20.9}$ & $78.1_{-29.0}^{+40.7}$ \\
\hline
\end{tabular}
\end{center}
\caption{Theoretical classical microlensing event rate (in units of $10^{-7}$ events/star/year) towards the centre of the selected globular clusters. For each model, we report the quantity $XX^{+YY}_{-ZZ}$ meaning that the estimated range for $\Gamma_{\rm c}$ is $(XX - ZZ, XX + YY)$ obtained by varying the slope $\alpha$ of the lensed MF in the interval $(\alpha_l, \alpha_u)$, while $XX$ is the value of $\Gamma_{\rm c}$ for the MF with slope $\alpha_0$. Both $(\alpha_l, \alpha_u)$ and $\alpha_0$ are given in Table\,1.}
\end{table*}

Table\,2 reports the optical depth towards the centre of the selected globular clusters and shows that $\tau$ is best suited to discriminate among different halo models since the values predicted are quite sensitive to both the flattening and the slope of the density profile. Furthermore, $\tau$ does not depend on the MF and hence possible degeneracies between the density parameters and the MF slope are avoided. Unfortunately, measuring the optical depth from pixel lensing events is not immediately possible. The usual estimator, $\tau_{\rm meas} = (\pi/2E) \sum{t_{E, i}/\varepsilon(t_{E, i})}$, may not be used since the Einstein time $t_E$ is not available from pixel lensing events. An alternative estimator has been proposed (\cite{G99}), but it turns out to depend on the luminosity function and also requires a detailed knowledge of the efficiency of the pixel lensing survey. Unfortunately, both these quantities are difficult to obtain or not available at all. 

Table\,3 gives the classical event rate towards the centre of the selected globular clusters. The dependence of $\Gamma_{\rm c}$ both on the halo models and the MF is clearly evident. Actually, an accurate estimate of $\Gamma_{\rm c}$ is not possible unless the uncertainty on the slope $\alpha$ of the lenses MF is significatively reduced. However, even if $\alpha$ were known with much less uncertainty, observations towards a single globular cluster can not uniquely discriminate among different halo models. For instance, Table\,3 shows that the two extreme models I50a and TFb predicts almost the same value (within the uncertainties) of $\Gamma_{\rm c}$ towards all clusters. It is, however, interesting to observe that the quantity $\Gamma_{\rm c}^{i}/\Gamma_{\rm c}^{tot}$, with $\Gamma_{\rm c}^{i}$ the rate towards the i\,-\,th cluster and $\Gamma_{\rm c}^{tot} = \sum_{i}{\Gamma_{\rm c}^{i}}$, is almost independent on the MF and can be used to discriminate among different halo models. The same results approximately applies if we use $\Gamma_{\rm pl}$ instead of $\Gamma_{\rm c}$. Since $\Gamma_{\rm pl}$ can be easily measured from observations, this could be an interesting method to discriminate among different halo models avoiding the uncertainties related to the lenses MF and the need for the measurement of $\tau$.

\begin{table}
\begin{center}
\begin{tabular}{|c|c|c|c|c|c|}
\hline
Name & $\mu_0, r_t, r_c$ & $\mu_V$ & $M_V$ & $Z$ & $\overline{\langle u_T \rangle}$ \\
\hline \hline
NGC 288 & 14.43, 12.94, 1.42 & 14.69 & -6.60 & 10 & 0.27 \\
NGC 1904 & 16.23, 8.34, 0.16 & 15.59 & -7.86 & 5 & 0.26 \\
NGC 2419 & 19.83, 8.74, 0.35 & 19.97 & -9.58 & 1 & 0.01 \\
NGC 4147 & 17.63, 6.31, 0.10 & 16.48 & -6.16 & 3 & 0.13 \\
NGC 5024 & 17.39, 21.75, 0.36 & 16.38 & -8.77 & 2 & 0.14 \\
NGC 5053 & 22.19, 13.57, 1.98 & 16.19 & -6.72 & 1 & 0.10 \\
NGC 5272 & 16.34, 38.19, 0.55 & 15.12 & -8.93 & 5 & 0.22 \\
NGC 5466 & 21.28, 34.24, 1.64 & 16.15 & -7.11 & 1 & 0.10 \\
NGC 5904 & 20.49, 12.07, 1.96 & 14.46 & -8.81 & 10 & 0.33 \\
NGC 6171 & 18.84, 17.44, 0.54 & 15.06 & -7.13 & 20 & 0.19 \\
NGC 6205 & 16.80, 25.18, 0.78 & 14.48 & -8.70 & 5 & 0.26 \\
NGC 6218 & 18.17, 17.60, 0.72 & 14.02 & -7.32 & 6 & 0.34 \\
NGC 6254 & 17.69, 21.48, 0.86 & 14.08 & -7.48 & 6 & 0.33 \\
NGC 6341 & 15.58, 15.17, 0.23 & 14.64 & -8.20 & 1 & 0.29 \\
NGC 6366 & 21.24, 15.20, 1.83 & 14.97 & -5.77 & 30 & 0.26 \\
NGC 7078 & 14.21, 21.50, 0.07 & 15.37 & -9.17 & 1 & 0.27 \\
NGC 7089 & 15.92, 21.45, 0.34 & 15.49 & -9.02 & 4 & 0.22 \\
Pal 13 & 23.36, 2.20, 0.48 & 17.31 & -3.51 & 4 & 0.05 \\
\hline
\end{tabular}
\end{center}
\caption{Main properties of the sample of globular clusters selected as described in the text. The King model parameters are the central surface brightness $\mu_0$ (in the V\,-\,mag/arcsec$^2$), the tidal radius $r_t$ and the core radius $r_c$ (in arcsec). The distance modulus $\mu_V$ and the integrated absolute visual magnitude $M_V$ fix the luminosity of the cluster. The last two columns gives the metallicity $Z$ (in units of $10^{-4}$) and the value of $\overline{\langle u_T \rangle}$.}
\end{table}

While $\tau$ and $\Gamma_{\rm c}$ do not depend on the target LF, this latter quantity plays a key role in the determination of $\langle u_T \rangle(x, y)$. In Table\,4, we report the main photometric parameters for the sample of globular clusters we are considering and the value of $\overline{\langle u_T \rangle}$ defined as follows\,:

\begin{equation}
\overline{\langle u_T \rangle} \equiv 
\frac{\int{\langle u_T \rangle(x, y) \times 10^{- \mu_{\rm GC}(x, y)/2.5} \ dx \ dy}}{\int{10^{- \mu_{\rm GC}(x, y)/2.5} \ dx \ dy}} 
\label{eq: defutover}
\end{equation}
which is the average threshold impact parameter spatially weighted with the cluster surface density. We will use this quantity later to approximatively estimate the number of observable lensing events.

\section{Conclusions}

The results in the previous section have shown that observations towards globular clusters have the potential to discriminate among halo models and to shed light on the MF of the lenses. Actually, in order to see whether a pixel lensing survey may indeed be useful, we must evaluate (at least qualitatively) the theoretical number of events predicted for a given model. We can estimate it as\,:

\begin{equation}
N_{\rm pl}^{ev} \simeq  N_{\star}(m \le m_{\rm lim}) \times T_{\\rm obs} \times \Gamma_{\rm c}(0, 0) \times 
\overline{\langle u_T \rangle}
\label{eq: nevpl}
\end{equation}
where $N_{\star}(m \le m_{\rm lim})$ is the number of cluster stars brighter than $m_{\rm lim}$ and $T_{\rm obs}$ the total period of observations, that is the product of the observability\footnote{The observability is strongly dependent on the telescope site. We have assumed that observations are performed using the TT1 telescope and deem a cluster as {\it observable} if it is higher than $20^o$ on the horizon for more than 3 hours per night.} per year and the duration of the monitoring campaign. In Eq.(\ref{eq: nevpl}), we have approximated\,:

\begin{displaymath}
\Gamma_{\rm c}(x ,y) \simeq \Gamma_{\rm c}(0, 0) \ \ , \ \ 
\langle u_T \rangle (x, y) \simeq \overline{\langle u_T \rangle} \ .
\end{displaymath}
While the first approximation works very well (within $2\%$), the second one leads to a slight overestimation of the number of events, but nonetheless is useful to get order of magnitudes results that are enough for our aims.

To estimate $N_{\star}(m \le m_{\rm lim})$, we have to choose a value for the limiting magnitude. This is a complicated task. In principle, all the stars in the cluster could be sources of a detectable pixel lensing event provided that the amplification is sufficiently high. Actually, it is highly unlikely that very faint stars give rise to detectable events. Furthermore, the limiting magnitude depends on the efficiency of the monitoring campaign and of the data analysis pipeline. We fix $m_{\rm lim} = 26$ for NGC\,7078 and scale $m_{\rm lim}$ to the other globular clusters according to the distance modulus. A higher value of $m_{\rm lim}$ should reduce the predicted number of events. To estimate $N_{\star}(m \le m_{\rm lim})$, we have first to renormalize each simulated LF using the condition\,:
 
\begin{equation}
C_{\rm norm} \ \int{10^{-M/2.5} \phi(M) dM} = 10^{-M_V/2.5}
\label{eq: normlf}
\end{equation} 
with $M_V$ the integrated absolute magnitude of the cluster. Then, the number of stars brighter than $m_{\rm lim}$ is simply\,:

\begin{equation}
N_{\star}(m \le m_{\rm lim}) = C_{\rm norm} \ \int_{M_{\rm min}}^{m_{\rm lim} - \mu_V}{\phi(M) dM}
\label{eq: nstar}
\end{equation} 
where $C_{\rm norm}$ is the normalization coefficient, $M_{\rm min}$ the minimum value of the absolute magnitude and $\mu_V$ the distance modulus. The total number of events predicted for pixel lensing observations is reported in Table\,5 for a monitoring campaign lasting three years. To study the impact of the metallicity of the cluster population on the estimated number of events, we arbitrarily assume that all the clusters of the sample have the same metallicity and use simulated luminosity functions for stellar populations having three different values of $Z$, as described before in Sect.\,3.  The observed trend with $Z$ is easy to explain qualitatively. The lower the metallicity, the fainter the stars in the same evolutionary phase and hence the higher the threshold amplification $A_T$ needed. Since $N_{\rm pl}^{\rm ev} \propto u_T \propto A_T^{-1}$, it is expected that $N_{\rm ev}^{\rm pl}$ decreases as the metallicity $Z$ increases. 

Unfortunately, whatever is the value of $Z$, it turns out that the number of expected events is so small that all the tests we have suggested to discriminate among halo models are useless because almost no events are predicted. It is important to remember that the results in Table\,5 have been obtained using Eq.(\ref{eq: gammapl}) which assumes a perfect detection efficiency and ideal observational conditions. Thus, the numbers in Table\,5 should be considered as strict upper limits, the actual number of detectable events being far lower.  

\begin{table}
\begin{center}
\begin{tabular}{|c|c|c|c|}
\hline
Model & $Z = 10^{-2}$ & $Z = 10^{-3}$ & $Z = 10^{-4}$ \\
\hline \hline
I50a & $0.73_{-0.40}^{+0.75}$ & $1.35_{-0.74}^{+1.37}$ & $1.70_{-0.92}^{+1.73}$ \\
I75a & $0.70_{-0.32}^{+0.55}$ & $1.29_{-0.58}^{+1.00}$ & $1.63_{-0.74}^{+1.26}$ \\ 
I100a & $0.66_{-0.31}^{+0.57}$ & $1.20_{-0.56}^{+1.06}$ & $1.52_{-0.71}^{+1.33}$ \\
TFa & $1.35_{-0.63}^{+1.19}$ & $2.49_{-1.16}^{+2.17}$ & $3.14_{-1.46}^{+2.75}$ \\
I50b & $0.49_{-0.18}^{+0.31}$ & $0.90_{-0.38}^{+0.53}$ & $1.13_{-0.48}^{+0.68}$ \\ 
I75b & $0.46_{-0.18}^{+0.26}$ & $0.86_{-0.34}^{+0.47}$ & $1.10_{-0.38}^{+0.59}$ \\ 
I100b & $0.43_{-0.16}^{+0.22}$ & $0.79_{-0.29}^{+0.41}$ & $1.00_{-0.37}^{+0.51}$ \\
ITFb & $0.89_{-0.33}^{+0.46}$ & $1.63_{-0.61}^{+0.85}$ & $2.06_{-0.77}^{+1.07}$ \\
\hline
\end{tabular}
\end{center}
\caption{Total number of events for the different models considered and the three simulated luminosity functions (with $Z$ the metallicity of the population). The central value refers to the MF with slope $\alpha_0$, while the lower and upper limits have been obtained varying the slope $\alpha$ in the range given for each model in Table\,1.}
\end{table}

Note also that, up to now, all the quantities reported have been given without any uncertainties. Actually, both the halo model parameters and the globular clusters properties (King model parameters and metallicity) are known with errors that should be propagated to obtain the uncertainties on the predicted microlensing quantities (optical depth, average threshold impact parameter, number of events). We have not considered these errors since they are far lower than the ones due to the MF uncertainties and the halo mass density profile. However, they should be correctly estimated before comparing the theoretical predictions with the (eventually obtained) observational results. On the other hand, having neglected these uncertainties does not affect our main results.

It is interesting to compare our results to those found by Rhoads \& Malhotra (1998) who concluded that a pixel lensing survey should be able to observe a total number of 40\,-\,120 events after 2\,-\,5 years of monitoring of the entire globular cluster system. The method used by these authors is quite different from ours since they estimate the number of events by multiplying the event rate by the {\it effective number of fully resolved stars}, defined as the integral of the cluster luminosity function weighted by a threshold impact parameter. This number is explicitly evaluated for NGC\,7078 and then scaled to the other clusters according to the integrated absolute magnitude and the distance modulus. Moreover, they assume that all the events have the same duration ($\simeq 30$ days), which is a very rough approximation, and that the halo is fully made out of MACHOs. If we scale our results to $f = 1$ and to the entire globular cluster system (i.e. we observe $\sim 150$ clusters instead of 18), we roughly get 11 - 130 events after three years of monitoring, which is in qualitatively good agreement with the results by Rhoads \& Malhotra (1998), especially if we consider that they include also the contribution of the bulge and the disk and use different halo models. This simple scaling argument may be considered as an evidence that our results are not significatively affected by systematics induced by the method we have used to get them. Actually, the smaller number of events with respect to that predicted by Rhoads \& Malhotra (1998) is a consequence of our most realistic assumptions on the halo modeling and of the lower number of targets. Thus we conclude that {\it pixel lensing observations towards galactic globular clusters are unable to discriminate among different halo models because of the too small number of detectable events.} 

While not useful to constrain the dark halo properties, globular clusters are still viable targets for a pixel lensing survey. In this paper, we have considered targets which are far above the galactic plane since we were mainly interested to the halo. However, the line of sight to many globular clusters goes through the disk and thus allows one to investigate its vertical structure and to extend the determination of its MF to the faint end, unreachable with optical observations. Furthermore, globular clusters near the galactic centre (such as, e.g., 47\,Tuc) may be used both as targets and to enhance the probability for lensing (\cite{Ursula}). Then, it should be interesting to extend the same computations of this paper to clusters located towards the centre of the Milky Way or towards spiral arms to see whether the number of events increases. 

The analysis performed in this paper has shown that globular clusters are not useful targets to investigate the dark halo structure. However, other possible targets are the dwarf satellites galaxies of the Milky Way (\cite{Lukas}). For many of them, the line of sight intercepts only the dark halo and hence the (eventually) observed microlensing events should be due to galactic MACHOs. Being dwarf galaxies dark matter dominated, it is expected that self lensing could seriously contribute to the number of events complicating the interpretation of the results. A detailed modeling of the target is thus needed in order to decouple the self lensing contribution from the Miky Way dark halo one. Even if difficult, this task is nowaday possible (see, e.g., \cite{KW02}, \cite{WK02} for Draco). In a forthcoming paper, we plan to repeat our analysis to see whether pixel lensing observations towards some dwarf satellite galaxies could shed light on the galactic dark halo. 

\begin{acknowledgements}

A preliminary version of this work was presented during an informal meeting of the SLOTT\,-\,AGAPE collaboration held in Zurich. We warmly thank all the partecipants to that meeting for the helpful discussions we had on the manuscript. We are also grateful to Martin Dominik for his suggestions about the selection criteria of the target globular clusters. W. E. Harris is acknowledged for having made publicy available his catalogue of globular clusters. Finally, we acknowledge the referee, S. Portegies Zwart, for comments which have helped to improve the presentation.

\end{acknowledgements}

\end{document}